\documentclass{aa}  

\usepackage{graphicx}
\usepackage{txfonts}

\usepackage{graphicx} 
\usepackage[fleqn]{amsmath}
\usepackage{array, amssymb}	
\usepackage[utf8]{inputenc}
\usepackage[T1]{fontenc}   
\usepackage{url}       
\usepackage{booktabs}    
\usepackage{lmodern}

\usepackage{hyperref}

\makeatletter
\newcommand*\mysize{%
  \@setfontsize\mysize{6.5}{9.0}%
}
\makeatother

\usepackage{xcolor}
\definecolor{keywordcol}{rgb}{0.00, 0.44, 0.13}
\definecolor{stringcol}{rgb}{0.73, 0.13, 0.13}
\definecolor{numbercolor}{rgb}{0.25, 0.63, 0.44}
\definecolor{identifiercolor}{rgb}{0.0, 0.0, 1.0}
\usepackage{listings}
\lstdefinestyle{mypython}{
    language=Python,
    frame=lines,
    basicstyle=\ttfamily\mysize, % adjust font size here
    keywordstyle=\color{keywordcol}\bfseries,
    commentstyle=\color{gray},
    numberstyle=\color{numbercolor},
    stringstyle=\color{stringcol},
    showstringspaces=false,
    breaklines=true,
    rulecolor=\color{black},
}

\usepackage{xspace}

\newcommand{\hessj}{HESS\,J1825$-$137\xspace}
\newcommand{\psr}{PSR\,B1823$-$13\xspace}
\newcommand{\code}[1]{\texttt{#1}}

\begin{document} 

   \title{Energy-dependent gamma-ray morphology estimation tool in Gammapy}

   \author{K. Feijen\thanks{Email: kirsty.feijen@gmail.com} \inst{1}
          \and R. Terrier \inst{1}
          \and B. Khélifi \inst{1}
          \and A. Sinha \inst{2}
          \and A. Donath \inst{3}
          \and A. Mitchell \inst{4}
          \and Q. Remy \inst{5}
          }

   \institute{Universit\'e Paris Cit\'e, CNRS, Astroparticule et Cosmologie, F-75013 Paris, France \\
    \and Tata Institute of Fundamental Research, 1 Homi Bhabha Road, Colaba, Mumbai, 400 005, India\\
    \and Center for Astrophysics | Harvard \& Smithsonian, 60 Garden Street, Cambridge, MA\\
    \and Friedrich-Alexander-Universit\"at Erlangen-N\"urnberg, Erlangen Centre for Astroparticle Physics, Nikolaus-Fiebiger-Str. 2, 91058 Erlangen, Germany\\
    \and Max-Planck-Institut f\"ur Kernphysik, P.O. Box 103980, D 69029 Heidelberg, Germany\\
             }

   \date{Received ''; accepted ''}

  \abstract
   {An understanding of the energy dependence of gamma-ray sources can yield important information on the underlying emission mechanisms. However, despite the detection of energy-dependent morphologies in many TeV sources, we lack a proper quantification of such measurements.}
   {We introduce an estimation tool within the Gammapy landscape, an open-source Python package for the analysis of gamma-ray data, for quantifying the energy-dependent morphology of a gamma-ray source.}
   {The proposed method  fits the spatial morphology in a global fit across all energy slices (null hypothesis) and compares this to separate fits for each energy slice (alternative hypothesis). These are modelled using forward-folding methods, and the significance of the variability is quantified by comparing the test statistics of the two hypotheses.}
   {We present a general tool for probing changes in the spatial morphology with energy, employing a full forward-folding approach with a 3D likelihood. We present its usage on a real dataset from H.E.S.S. and on a simulated dataset to quantify the significance of the energy dependence for sources of different sizes. In the first example, which utilises a subset of data from \hessj, we observe extended emission at lower energies that becomes more compact at higher energies. The tool indicates a very significant variability ($9.8\sigma$) in the case of the largely extended emission. In the second example, a source with a smaller extent ($\sim0.1^{\circ}$), simulated using the CTAO response, shows the tool can still provide a statistically significant variation ($9.7\sigma$) on small scales.}
   {}

   \keywords{VHE Gamma-ray phenomena -- Analysis techniques -- Energy-dependent propagation}

   \maketitle

\section{Introduction}

\label{sec:intro}
Detecting variations in the morphology of gamma-ray sources with energy can offer useful insight into their underlying physical mechanisms. However, measuring and quantifying these variations remains difficult. These energy-dependent morphologies are expected in a variety of astrophysical sources, including pulsar wind nebulae (PWNe), supernova remnants (SNRs), and relativistic jets. It can therefore be helpful to look for any evidence of this energy dependence.

The evolutionary phases of a PWN lead to a unique gamma-ray morphology \citep{Gaensler_2006}. The emission is expected to be compact and close to the pulsar position at high energies. However, for lower energies, the emission is expected to be more extended, with an offset from the pulsar position. Radiative cooling, either via synchrotron emission in magnetic fields or the inverse Compton scatter of ambient photon fields, leads to higher energy electrons losing energy faster. This creates a spectral steepening with distance. Another physical process at play is the energy-dependent diffusion of particles, in which high-energy particles diffuse more efficiently than the lower energy ones. An energy-dependent morphology can also be attributed to the bulk motion of particles in the PWN outflow. Advection-dominated transport in this case leads to a spatial distribution that is dependent on the energy loss timescale.
Conversely, for SNRs with uniform gas densities, we expect the gamma-ray emission at lower energies to be concentrated to the central SNR position and the high-energy emission to be farther away \citep{2019_edep_SNR,2021_edep_SNR}. 
Relativistic jets, such as those observed in microquasars, are also expected to exhibit an energy-dependent morphology along their extent \citep{SS433_Laura_2024}.

The High Energy Stereoscopic System (H.E.S.S.) has detected several sources exhibiting energy-dependent morphologies. Examples include evolved PWNe such as HESS\,J1303$-$631 \citep{Edep_J1303_2012} and \hessj, where the nebula size increases with decreasing energy. \hessj is a well-known example of a PWN that shows a clear energy-dependent gamma-ray morphology \citep{Edep_PWNe_HESS_2006,Edep_PWNe_HESS_2019,Edep_Principe_2020}. A compact nebula with an extended diffuse nebula and an asymmetric morphology was discovered in X-rays towards \psr  \citep{Finley_1996}. \cite{HESS_2005} discovered an extended very high-energy (VHE) gamma-ray nebula towards this region, designated \hessj. The size of the nebula is seen to increase with decreasing energy \citep{Edep_PWNe_HESS_2019}. At low energies ($< 32$\,TeV) the emission nebula is extended, with its peak offset from the position of \psr. For higher energies ($> 32$\,TeV), the emission is more compact and close to the pulsar position. This behaviour indicates that the gamma-ray emission towards \hessj originates from PSR\,B1823$-$13 as a PWN. 

SS\,433 was the first microquasar to be detected at TeV energies \citep{HAWC_micro_2018,SS433_Laura_2024}. The H.E.S.S. gamma-ray image, with the nearby extended emission from HESS\,J1908$+$063 modelled out, reveals two regions of emission, known as the eastern and western jets. No TeV emission is significantly detected from the central source. At the highest energies (>10 TeV), the emission appears only in the base of the X-ray jets, i.e. the point closest to the central source. In contrast, the lower energy gamma rays (0.8–2.5 TeV and 2.5–10 TeV) peak in significance at positions farther along the jets. 

Various techniques, applicable in different scenarios, have been utilised to search for energy dependence. These include the comparison of spectral indices across different regions or energy bins in a map \citep{Edep_PWNe_HESS_2006}, slicing spatial maps into different energy bands and comparing their extension to a model fit (see examples of Gaussian spatial model fits in \citealt{Edep_J1303_2012}, \citealt{Edep_Principe_2020}, and \citealt{Edep_MAGIC_2020}), comparing radial profiles \citep{Edep_Principe_2020}, and many other techniques. 
Energy-dependent morphology is a key observable for studying the nature of a VHE gamma-ray source. 
Here, we present a new tool in the Gammapy framework \citep{gammapy_ref,gammapy_zenodo_12} for investigating the potential energy-dependent morphology towards a gamma-ray source. 
This method utilises maps across multiple energy bands to fit the morphological parameters of the source, such as position and extension. 
The user defines the best-fit model for the morphology and spectrum. 
A global fitting approach is applied, and then an individual fit is performed in each energy band to assess if the morphology changes significantly.

\section{Methodology}
\label{sec:methods}
In this section we describe various methods for measuring the energy dependence of gamma-ray sources, highlighting their respective advantages and disadvantages. 

One common approach is spatially resolved spectral analysis \citep{Aharonian_2006,2006_RXJ}. This technique can be used to search for a softening of the gamma-ray emission away from the pulsar position in a PWN scenario, which provides the evidence of an energy-dependent morphology.
The signal region is divided into non-overlapping subregions. A spectral model, typically a power law, is fit in each subregion independently to search for changing photon index across the source region \citep{2006_RXJ}. This approach offers a simple method of analysing the spectra, particularly for variations in spectral shape. However, this method has its limitations. The choice of region size and number of regions can strongly influence the result, leading to potential biases. The astrophysical background is difficult to determine accurately for each region, the estimation for different regions are typically not independent. These limitations are more significant for small subregions that approach the resolution of the point spread function (PSF), making this method better suited to more extended sources. Quantifying the significance of spectral variability across these subregions remains a challenge.

Another method involves creating radial excess profiles along a specific axis \citep{HESS_2012,Edep_PWNe_HESS_2019,Edep_Principe_2020}. The distribution of the emission is analysed by extracting the excess counts from specific regions in a specific orientation from the source position. This is useful to visualise changing spatial features such as extension and intensity. 
It is important to note that they do not take into account the PSF or variations in exposure. 
Radial profiles offer a model-independent method for assessing the spatial variations, allowing for an unbiased way of visualising the data. However, they assume radial symmetry within the region, which may not always be a valid approximation. Additionally, this method is highly sensitive to the selected axis direction of the profile and the choice of the radial bins. Another limitation is the difficulty in handling background systematics. Consequently, these dependencies make it a less robust analysis. 
Additionally, variations in the exposure across the field of view can change the excess distribution, meaning the excess partially reflects these exposure variations rather than the source properties itself. For example, a decrease in excess counts may be misinterpreted as a physical or temporal effect, when it is, in fact, stemming from the exposure variation. This makes deriving the true significance of the effect a challenge. 
While traditionally overlooked, accounting for exposure variation is now possible with the introduction of methods like the \code{FluxProfileEstimator} in Gammapy. This, however, still does not account for contributions from additional sources within the field of view or analysing variations on small spatial scales.

The final method discussed here involves slicing the spatial maps in different energy bins and estimating the extension of the source \citep{Edep_Principe_2020} by taking into account all instrument response functions (IRFs), then deconvolving from both instrumental and exposure effects. In this case, a model is utilised to fit the morphological parameters of the source in these energy bins, providing valuable insight into the spatial evolution of the emission. While this approach allows for a detailed fitting, treating the fitted parameters independently can lead to an underestimation of the trial factors, thereby reducing the reliability of the results. Along with the other limitations discussed previously, this shows the need for a more robust global testing method. 
In some cases, it may be possible to directly fit a custom parametric model with energy-dependent position and size. However, such models require an initial understanding of the expected changes in the spatial parameters.

We propose a new method that addresses each of the limitations described above, to quantify the energy-dependent morphology with a proper statistical assessment. This technique utilises a full forward-folding approach, where a morphological fit is conducted across various energy slices and compared with a global morphology fit. The technique chosen here involves fitting a spatial model and comparing the differences in morphological parameters (such as extension and position) in different energy bins. 
Further information about the tool and examples can be found in Sects. \ref{sec:description} and \ref{sec:examples}, respectively.

In gamma-ray astronomy we often utilise hypothesis testing. We rely on Wilks' theorem \citep{Wilks}, which is a log-likelihood ratio test that allows us to estimate the relative significance through the test statistic ($\Delta\rm{TS}$),

\begin{equation}
\Delta\rm{TS} = -2 \ln \lambda = -2 \ln \left( \dfrac{\mathcal{L}_0^{\rm{max}}}{\mathcal{L}_1^{\rm{max}}} \right) \, ,
\label{eqn:delta_ts}
\end{equation}where $\mathcal{L}_0^{\rm{max}}$ and $\mathcal{L}_1^{\rm{max}}$ represent the maximum likelihood of the model under the null and alternative hypotheses, respectively. 
Under the correct conditions, where the two models are nested, and the regularity conditions are satisfied, we expect the distribution of $\Delta\rm{TS}$ to asymptotically approach a $\chi^2$ function with $m$ degrees of freedom, the difference in number of free parameters between the two nested hypotheses. 
Therefore, for a large enough sample it is possible to compute the $\Delta\rm{TS}$ and compare to a chi-squared function with the correct degrees of freedom to test the statistical significance. 

For our tool, the null hypothesis (H$_0$) is defined by a global fit. Initially, the morphological parameters are fit in the first energy band. The resulting parameters are then applied to all subsequent energy bands and a fit is performed across the different energy bands. The alternative hypothesis (H$_1$) is defined by individual fits. Here, the dataset is divided into energy bins, and both the spatial and spectral models are fitted independently in each energy band. We apply Wilks' theorem to these hypotheses to obtain the $\Delta\rm{TS}$ (see Eq. \ref{eqn:delta_ts}). The $\Delta\rm{TS}$ is converted into a significance value through the survival function of the chi-squared distribution. 
Thus, this method provides a quantitative measure of how much the morphology deviates from the null model for each energy band.

\section{Implementation}
\label{sec:description}
The above algorithm was introduced in Gammapy v1.2 \citep{gammapy_ref, gammapy_zenodo_12} in a dedicated class called the \code{EnergyDependentMorphologyEstimator}. Users begin by using a previously optimised model on the relevant source of interest. This estimator runs on a \code{MapDataset} object, which combines the IRFs, counts, background and models in the region of interest.
Representing a 3D cube with two spatial dimensions and an energy axis, this dataset enables both spectral and morphological fitting, which is of interest here.

To test for energy-dependent morphology using our tool, one begins by dividing the \code{MapDataset} into different energy bins, as specified by the user. Firstly, we allow (at a minimum) the position and extension of the \code{SpatialModel} to be free, along with the amplitude of the \code{SpectralModel} and normalisation of the background. 
It is up to the user to define which spatial parameters are to be kept free. 
For example, for a \code{GaussianSpatialModel} there are five free parameters --- longitude, latitude, sigma, phi, and e. It is up to the user to make a judicious choice to freeze or free these parameters. In this case, longitude, latitude, and sigma are kept free.
For a proper energy-dependent investigation, it is necessary to keep at least the position and extension free. This enables us to detect any change in the position or size of the gamma-ray source across different energy bands. 
The spectral index can be poorly constrained, particularly when fitting in small energy bands. Therefore, it is recommended to assume a given spectral shape while allowing the amplitude to be fit. This re-normalisation can account for any small variations in the spectrum. 

Whilst there are several \code{SpatialModel}s available in Gammapy, it is necessary to select a model that accurately represents the morphology of the gamma-ray source. \code{SpatialModel}s such as the \code{DiskSpatialModel} and \code{ShellSpatialModel}, assume sharp edges, which are often not a good representation of the data, particularly across multiple energy bands. 
In contrast, the \code{GaussianSpatialModel} is able to represent the smooth features observed in gamma-ray morphology, particularly when the morphology varies with energy. This makes it a robust default for many analyses.
Alternative models should be used with caution and only when clearly justified by the data.
A usage example is shown in Fig. \ref{fig:code_snippet}, corresponding to the example in Sect. \ref{subsec:example2}.

\begin{figure}[!htb]
\begin{lstlisting}[style=mypython]
from astropy import units as u
from astropy.coordinates import SkyCoord
from IPython.display import display
from gammapy.datasets import MapDataset, Datasets
from gammapy.modeling.models import (
    GaussianSpatialModel,
    PowerLawSpectralModel,
    SkyModel,
)
from gammapy.estimators import EnergyDependentMorphologyEstimator

dataset = MapDataset.read(
    "dataset_energy_dependent.fits.gz"
)
datasets = Datasets([dataset])
energy_edges = [0.3, 2, 10, 100] * u.TeV

src_pos = SkyCoord(21.5, -0.89, unit="deg", frame="galactic")
spectral_model = PowerLawSpectralModel(
    index=2.4, 
    amplitude=4.22e-13 * u.Unit("cm-2 s-1 TeV-1"), 
    reference=1.0 * u.TeV,
)
spatial_model = GaussianSpatialModel(
    lon_0=src_pos.l,
    lat_0=src_pos.b,
    frame="galactic",
    sigma=0.2 * u.deg,
)

# Limit the search for the position on the spatial model
spatial_model.lon_0.min = src_pos.galactic.l.deg - 0.8
spatial_model.lon_0.max = src_pos.galactic.l.deg + 0.8
spatial_model.lat_0.min = src_pos.galactic.b.deg - 0.8
spatial_model.lat_0.max = src_pos.galactic.b.deg + 0.8

model = SkyModel(
    spatial_model=spatial_model, 
    spectral_model=spectral_model, 
    name="src",
)
model.spectral_model.index.frozen = True
datasets.models = model

estimator = EnergyDependentMorphologyEstimator(
    energy_edges=energy_edges, 
    source="src",
)

results = estimator.run(datasets)
results_edep = results["energy_dependence"]
display(results_edep["result"])
print(f"df = {results_edep['df']}")
print(f"deltaTS = {results_edep['delta_ts']}")
\end{lstlisting}
\caption{Using the \code{EnergyDependentMorphologyEstimator} object from \code{gammapy.estimators} to quantify any energy-dependent morphology in the source of interest. The \code{energy\_edges} defined are used to investigate the potential of energy-dependent morphology in the dataset.
The \code{gammapy.modeling.models} is used to define a source model with a spectral and spatial component. The initial parameter values are taken from a previously optimised model based on the full dataset, which accurately represents the source's morphology and spectrum. The position, extension and amplitude are let to be free as described in the text. The output of the code example is shown in \autoref{fig:code_output}.}
\label{fig:code_snippet}
\end{figure}

\section{Application examples}
\label{sec:examples}
\subsection{Example 1: \hessj}
\begin{figure}[!htb]
\centering
\includegraphics[width=0.99\columnwidth]{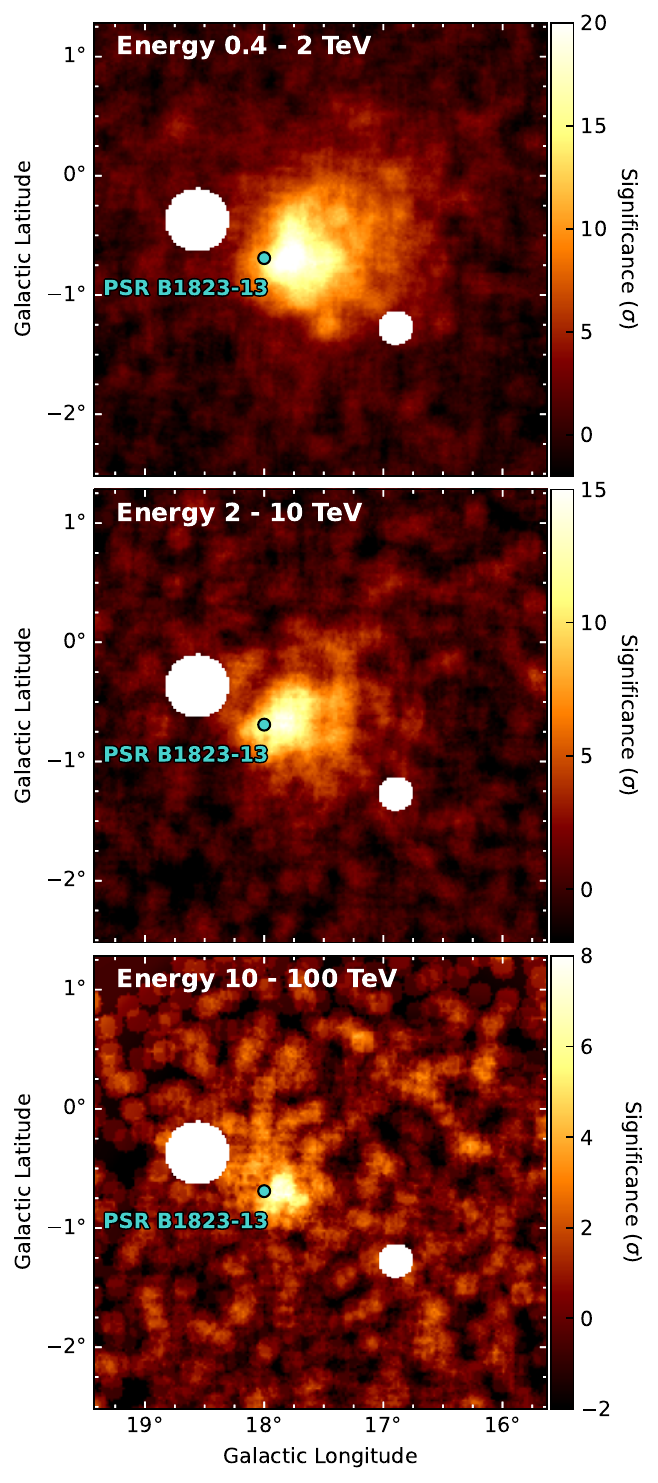}
\caption{Significance map of the H.E.S.S. dataset towards \hessj in the various energy bins of interest. An oversampling radius of $0.1^{\circ}$ is used. The white circles show the mask applied to the emission microquasar LS\,5039 and the gamma-ray source HESS\,J1826$-$130. The turquoise dot indicates the position of the pulsar, \psr. There is a clear change in the size of the emission in the different bins.}
\label{fig:dataset_ebands}
\end{figure}

To demonstrate the functionality of the tool, we showcase two examples. In the first example, we utilise a subset of the H.E.S.S. data towards \hessj from \cite{Aharonian_2006}. Three energy bins are chosen for this study: 0.4-2\,TeV, 2-10\,TeV and 10-100\,TeV. For this specific example, a mask is applied to exclude the emission from the microquasar LS\,5039 and the source HESS\,J1826$-$130. As shown by Fig. \ref{fig:dataset_ebands}, the dataset shows three distinct energy bins. 

The \code{PowerLawSpectralModel} and \code{GaussianSpatialModel} provide reasonable initial spectral and spatial models, respectively, for the dataset, so we used them for our study. 
The power law spectral shape is defined as follows:

\begin{equation}
\phi(E) = \phi_0 \cdot \left( \frac{E}{E_0} \right)^{-\Gamma}
\label{eqn:pwl}
, \end{equation}where $\phi_0$ is the amplitude, $\Gamma$ is the index of the power law fit, and $E_0$ is a normalisation energy taken here to be 1\,TeV. 
We utilise the \code{GaussianSpatialModel} in Gammapy, keeping the asymmetry frozen to 0 for simplicity, in the different energy bands, to test for any energy-dependent morphology with our tool. 

An important aspect of this new tool is to check that the energy bands chosen contain significant data. To do this, we can look at the signal above the background in each energy bin. This done by slicing the dataset into its energy bins. The null hypothesis (H$_0$) involves freezing all parameters except for the normalisation of the background, which is fitted in each energy band. The alternative hypothesis (H$_1$) allows additional parameters to be free, in this case the position and extension of the spatial model and the amplitude of the spectral model. Subsequently, the fit is performed for each energy band. These hypotheses are compared using Wilks' theorem (Eq. \ref{eqn:delta_ts}) to obtain a $\Delta\rm{TS}$ that indicates the significance of the signal relative to the background in each energy band. If each energy bin is significant above the background ($\gg5\sigma$), the energy-dependent study can continue. 

\begin{figure}[!htb]
\centering
\includegraphics[width=0.98\columnwidth]{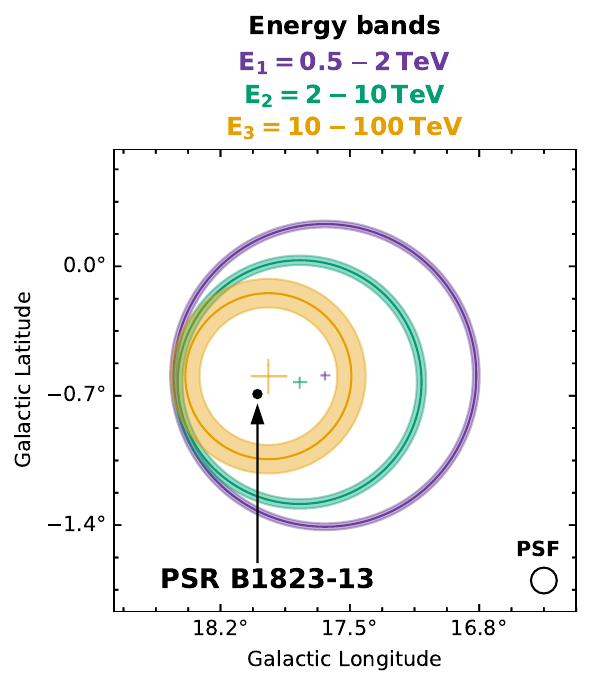}
\caption{Results of the estimator tool for the \hessj dataset. The different coloured rings show the fitted \code{GaussianSpatialModel} in the various energy bins along with their uncertainties. The plus symbols show the position and error from the fit. The black dot indicates the position of the pulsar, \psr.}
\label{fig:spatial_ebands}
\end{figure}

The next step is to quantify any energy-dependent morphology. The null hypothesis is determined by applying the same set of parameters to all energy bands and performing the fit. In the alternative hypothesis, the free parameters of the model are fit individually within each energy band. 
We plotted the fitted spatial models from our energy-dependent study. These are shown for each energy band in Fig. \ref{fig:spatial_ebands}.  

The $\Delta$TS values for each energy bin from the individual fit (H$_1$) are summed and subtracted from the $\Delta$TS of the global fit (H$_0$) to obtain a final $\Delta$TS value. This, along with the number of degrees of freedom, can be used to quantify the significance through the chi-squared test. In this example, the significance is 9.8$\sigma$, indicating that this source exhibits strong energy-dependent morphology, as expected from previous studies \citep{Aharonian_2006,Edep_Principe_2020,Edep_PWNe_HESS_2019}. We see at lower energies the emission is large and offset from the pulsar position, whilst at higher energies the emission is more compact and close to the pulsar position. 

\cite{Aharonian_2006} investigated the spectral properties of \hessj in different regions to search for any energy-dependent morphology. A fit to these results gives a chi-squared value of 58 with 11 degrees of freedom. This shows that the method outlined in our paper is more sensitive.

\subsection{Example 2: Slightly extended source}
\label{subsec:example2}
The second example focuses on a younger, less evolved object, which has had less time for electron cooling to occur, resulting in a smaller size. This object is expected to exhibit similar morphology to the previous case but on a much less extended scale; therefore, all traditional methods are ineffective. 
For this weakly extended PWN, there is ambiguity about whether the source size remains consistent across different energy bins at the scale of the PSF ($0.1^{\circ}$). 
Due to the small scale, it is not feasible to utilise other energy-dependent methods, as the source size approaches the level of the PSF. For example, to create a spatially resolved map of spectral indices would require an insufficient size for the boxes to obtain accurate results.

Based on the above information, we created a mock dataset that exhibits energy-dependent morphology. This consisted of simulating a 3D dataset with a live time of 50\,hours based on the Cherenkov Telescope Array Observatory (CTAO) IRFs \citep{CTAO_IRFs}, assuming a specific source model. 
The source is slightly extended as described by a circular emission region that follows a power law distribution $\sigma=\sigma_0 E^{-\alpha}$, where $\sigma_0=0.1^{\circ}$, $\alpha=0.3$ and $E$ is the energy axis for the given geometry.
The source has a flux of a few percent of the Crab flux at 1\,TeV ($4.22\times 10^{-13}$\,cm$^{-2}$\,s$^{-1}$\,TeV$^{-1}$) and an index of 2.42, mimicking the gamma-ray source HESS\,J1833$-$105 \citep[a composite SNR and PWN system,][]{hgps_2018,J1833_2008}. These models are assigned to the dataset to predict the counts data based on a Poisson probability distribution. 
To test for the energy-dependent morphology using the tool, we chose three energy bins for this example: 0.2-2\,TeV, 2-10\,TeV and 10-100\,TeV. 

\begin{figure}[htbp]
\centering
\includegraphics[width=0.98\columnwidth]{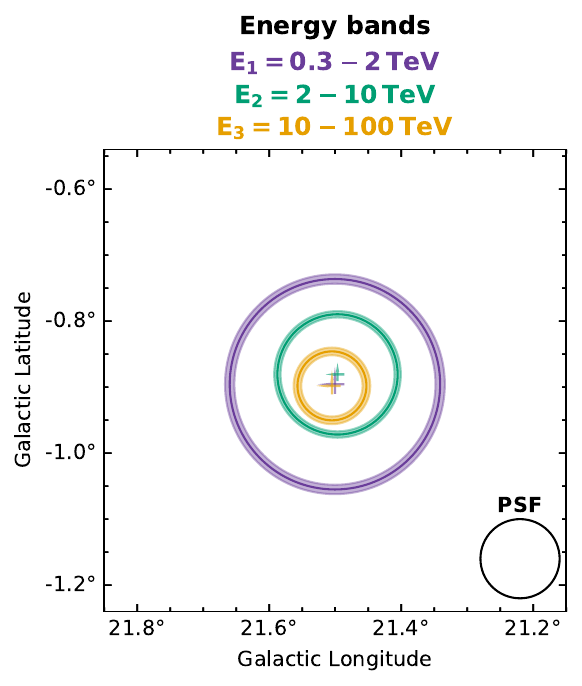}
\caption{Results of the estimator tool for the mock dataset. The different coloured rings show the fitted \code{GaussianSpatialModel} in the various energy bins along with their uncertainties. The plus symbols show the position and error from the fit.}
\label{fig:spatial_example2_ebands}
\end{figure}

As shown in Fig. \ref{fig:spatial_example2_ebands}, it is possible to distinguish the morphology in different energy bins at small scales ($<0.1^{\circ}$). The energy-dependent morphology estimator yields a significant result of 9.7$\sigma$. It is shown that at low energies the emission is significantly more extended than at higher energies.

\begin{figure}[htbp]
\centering
\includegraphics[width=0.98\columnwidth]{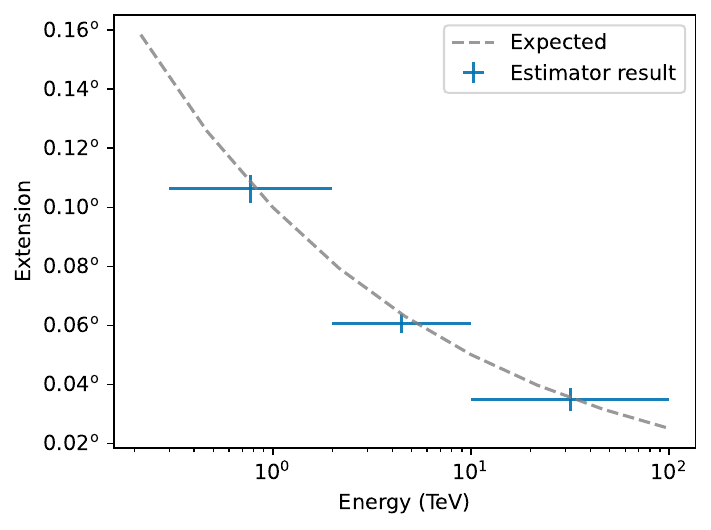}
\caption{Comparison of the variation in the extension with energy. The results obtained from the estimator for example 2 are shown in blue and the expected extension values taken from the simulated source sizes in grey. The grey line is described by a power law of $\sigma_0 E^{-\alpha}$, where $\sigma_0=0.1^{\circ}$ and $\alpha=0.3$.}
\label{fig:spatial_example2_curve}
\end{figure}

It is important to verify that the output of the energy-dependent estimator tool is consistent with the input parameters of the simulated source. The simulated source morphology is described by a power law distribution, as described previously. Figure \ref{fig:spatial_example2_curve} shows the estimator results in blue as compared to the expected simulated source sizes in grey, indicating good agreement.

\section{Validations}
\begin{figure}[htbp]
\centering
\includegraphics[width=0.98\columnwidth]{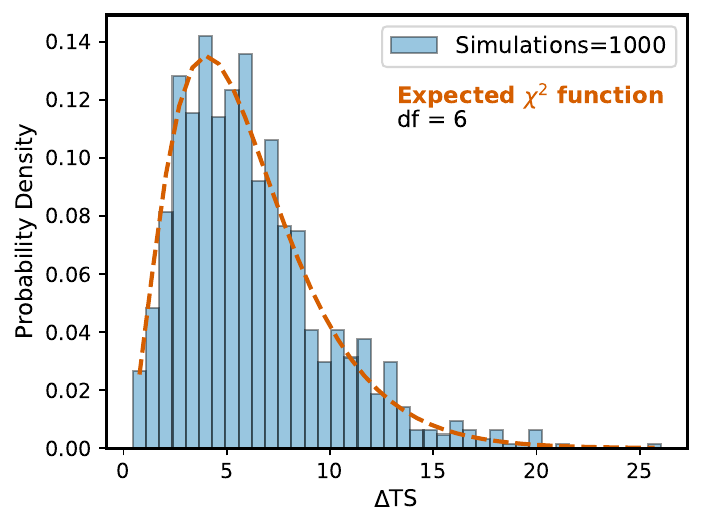}
\caption{Distribution of the samples of $\Delta$TS, calculated through Eq. \ref{eqn:delta_ts}, for each of the 1000 simulations. The expected chi-squared function has six degrees of freedom, which is the difference between the number of free parameters in the global fit (H$_0$) versus the individual fits (H$_1$).}
\label{fig:delta_TS}
\end{figure}

As previously mentioned, Wilks' theorem \citep{Wilks} states that the $\Delta$TS follows a chi-squared distribution for a large number of samples. To validate the accuracy of our method, we needed to conduct tests to ensure this holds true. We utilised a bootstrap parametric method in which the same method as the energy-dependent tool was used. 
We generated a simulated Poisson distributed dataset based on the model of interest for the chosen source. This mock dataset consisted of 10\,hours of observations based on the CTAO IRFs \citep{CTAO_IRFs} and exhibits a clear energy-dependent morphology. 
This simulation was performed 1000 times, resulting in a large sample set that contains the various parameters of interest including $\Delta$TS, as well as the latitude, longitude, and sigma for each sample. Figure \ref{fig:delta_TS} shows how the $\Delta\rm{TS}$ is distributed compared to the expected chi-squared function, with six degrees of freedom corresponding to the difference in the number of free parameters between the null and alternative hypotheses, as described in \autoref{sec:methods}.

The significance of the results from the energy-dependent estimator tool can also be expressed as the chi-square values for each parameter, $p$:
\begin{equation}
\chi_p^2 = \sum_i \dfrac{(x_i - \hat{\mu})^2}{\sigma^2}  \, ,
\label{eqn:chi2}
\end{equation}where $\hat{\mu}$ is the weighted mean (weighted by $1/\sigma_i$) and $x_i$ indicates the parameter value for each energy bin. It is important to note that the chi-squared parameter does not include potential correlation between the parameters, so it should be utilised cautiously. 

\begin{figure}[htbp]
\centering
\includegraphics[width=0.98\columnwidth]{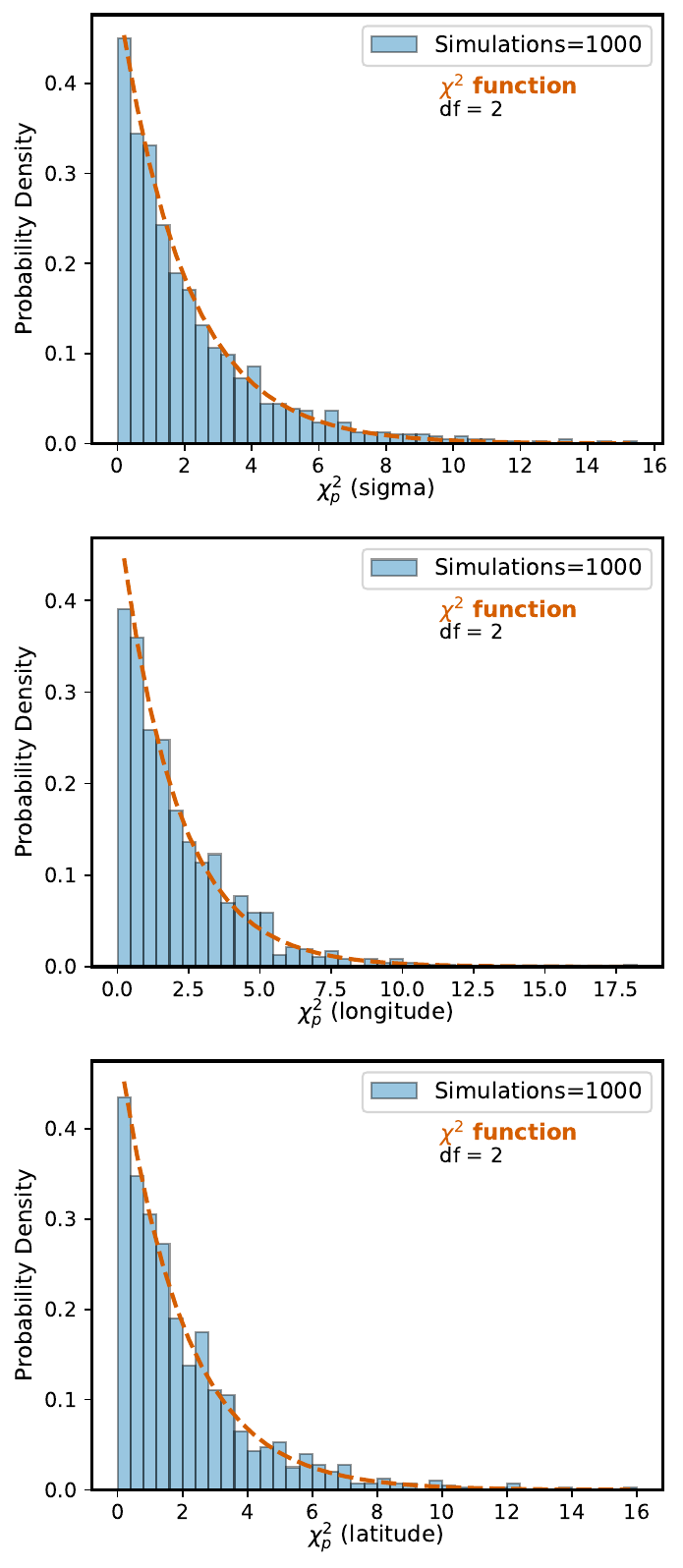}
\caption{Distribution of $\chi_p^2$ values as calculated through Eq. \ref{eqn:chi2} for each of the free parameters in the energy-dependent estimator tool.}
\label{fig:chi2}
\end{figure}

Figure \ref{fig:chi2} shows the distribution of the weighted chi-squared values for each of the free spatial parameters (longitude, latitude and sigma) calculated via Eq. \ref{eqn:chi2} for each sample set. The results follow a $\chi^2$ distribution, with the number of degrees of freedom corresponding to the number of energy bands minus one.

The distributions in Figs. \ref{fig:delta_TS} and \ref{fig:chi2} show good agreement with Eq. \ref{eqn:chi2}, especially where the sample density is highest. The deviations between the simulated distribution and the corresponding line equation, especially in the tail regions, are likely attributed to having an insufficient number of samples. Insufficient sampling can also lead to statistical fluctuations, resulting in decreased precision therefore making it harder for the simulated data to perfectly match the theoretical expectation. 
To test the compatibility of these we utilised a Kolmogorov-Smirnov test for goodness of fit. Typically a threshold p-value of 0.05 is utilised to indicate the data are compatible with the expected function. 
Comparison between the $\Delta$TS and expected chi-squared function in Fig. \ref{fig:delta_TS} leads to a p-value=0.097, supporting the notion that there is no significant difference between the two.
We perform the same test for Fig. \ref{fig:chi2} on the $\chi_p^2$ values, assuming they follow the expected chi-squared function with two degrees of freedom, and find a p-value$>0.05$ for each, providing evidence that the two are compatible.

\begin{figure}[htbp]
\centering
\includegraphics[width=0.98\columnwidth]{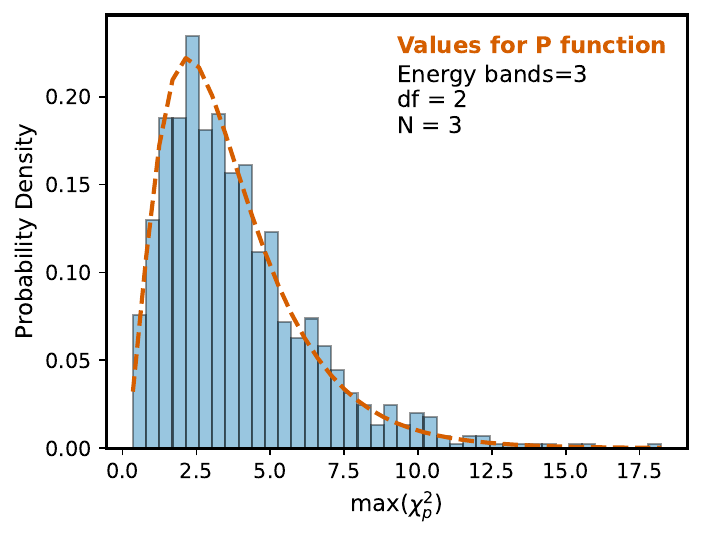}
\caption{Distribution of the $\max(\chi_p^2)$ value (longitude, latitude, or sigma) for each simulation.}
\label{fig:chi2_maximum}
\end{figure}

The $\chi_p^2$ values for each parameter (as shown in Fig. \ref{fig:chi2}) are calculated using Eq. \ref{eqn:chi2}. For each sample set, the three $\chi_p^2$ values (for sigma, longitude and latitude) are compared and the largest value is taken as the `$\max(\chi_p^2)$'. It is expected that these maximum values follow a specific distribution function:

\begin{equation}
P = N \, f(x, df) (F(x,df))^{N-1}
\label{eqn:P_value_chi2}
,\end{equation}where $N$ is the number of parameters, $f(x, df)$ is the probability density function (PDF) and $F(x,df)$ is the cumulative density function (CDF). Here, $x$ are the $\max(\chi_p^2)$ values and $df$ are the number of energy bands minus one. 
The derivation of Eq. \ref{eqn:P_value_chi2} is shown in Sect. \ref{sec:calc_chi2_max}. The distribution of the $\max(\chi_p^2)$ values, along with the expected function (Eq. \ref{eqn:P_value_chi2}) are shown in Fig. \ref{fig:chi2_maximum}.

\begin{figure}[!htb]
\centering
\includegraphics[width=0.98\columnwidth]{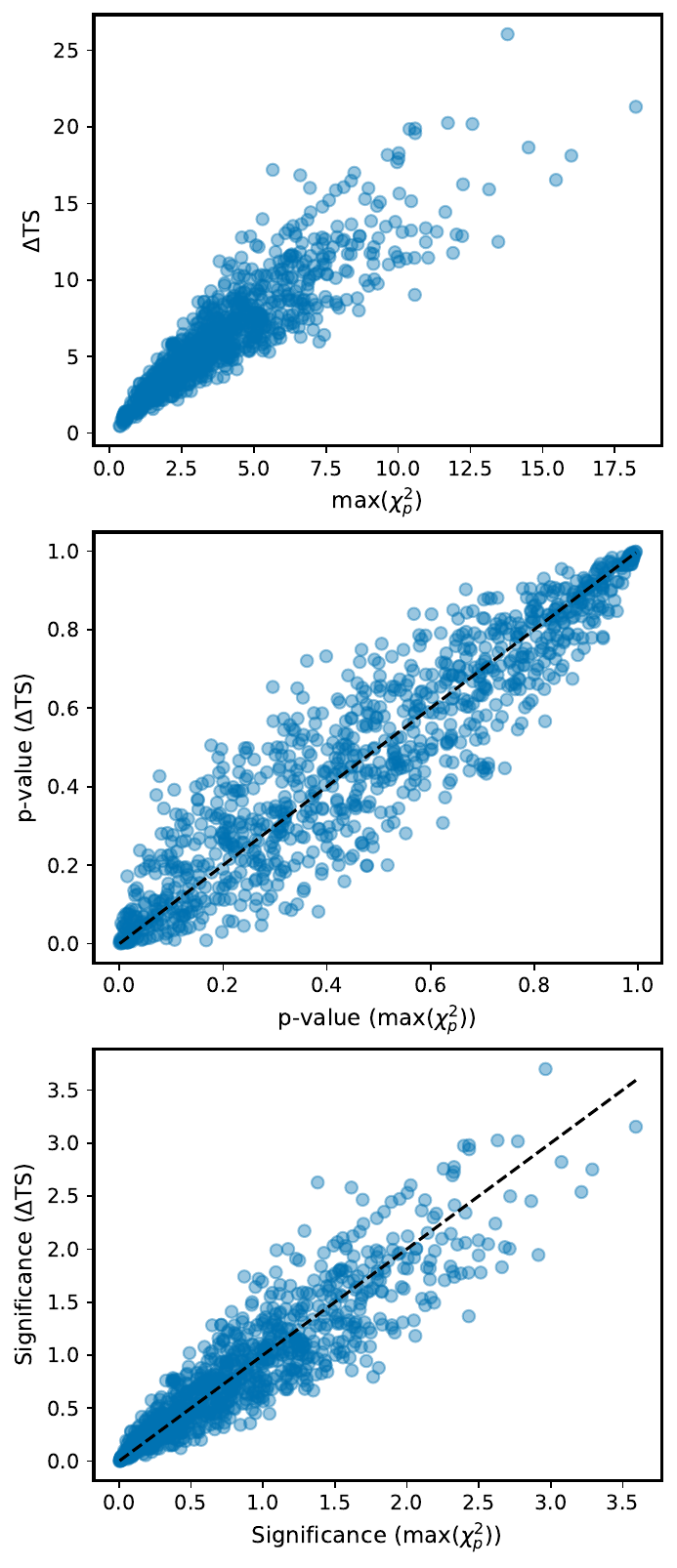}
\caption{Comparison of the $\Delta$TS and $\chi^2$ methods. \textit{Top}: Direct comparison of $\Delta$TS (Eq. \ref{eqn:delta_ts}) with the $\max(\chi_p^2)$ value (Eq. \ref{eqn:chi2}) from the three parameters (longitude, latitude, or sigma) for each simulation. \textit{Middle}: p-values for the $\Delta$TS and $\max(\chi_p^2)$ method as calculated through Eqs. \ref{eqn:pvalue_TS} and \ref{eqn:pvalue_chi2}, respectively. \textit{Bottom}: Significance for the $\Delta$TS and $\max(\chi_p^2)$ method as calculated through Eqs. \ref{eqn:sig_TS} and \ref{eqn:sig_chi2}, respectively. The dashed black line in the middle and bottom panels shows the line of unity ($x=y$). }
\label{fig:sensitivity2}
\end{figure}

The next step involves the comparison of the two distinct methods for calculating the significance, namely $\Delta$TS and $\chi^2$. This allows us to determine the compatibility of the results obtained from the two methods. 
As both the $\Delta$TS and max$(\chi_p^2)$ functions follow a $\chi^2$ distribution, we can directly compare their relationship on a scatter plot, as presented in the top panel of Fig. \ref{fig:sensitivity2}. 
It is worth reiterating that the $\Delta$TS method is the recommended approach, as it accounts for correlations between parameters, in contrast to the $\chi_p^2$ statistic, which treats each parameter independently. As a result, we do not necessarily expect a one-to-one relationship between the two.
We can also compare the p-values derived from each method. 
The equation for the p-value corresponding to the $\max(\chi_p^2)$ values is defined as follows:

\begin{equation}
p_{\rm{value}, \chi^2} = 1 - F(\max(\chi_p^2), df)^N \, ,
\label{eqn:pvalue_chi2}
\end{equation}where $F$ is the CDF that gives the probability of the random variable $X$ being less than or equal to $x$. The number of degrees of freedom, $df$, is equal to the number of energy bands minus one and $N$ is the number of parameters of interest, in this case equal to three. 

The p-value for the $\Delta$TS is calculated through\begin{equation}
    p_{\rm{value}, \Delta\rm{TS}} = s(\Delta\mathrm{TS}, df)
\label{eqn:pvalue_TS}
,\end{equation}where $s$ is the survival function (also defined as $1 - F$, where $F$ denotes the CDF). The number of degrees of freedom from the $\Delta$TS calculations is given by $df$, which in this case is equal to six. 
The comparison of the p-value from these methods are shown in the middle panel of Fig. \ref{fig:sensitivity2}. 

Finally, we can also determine the expected significance from each of the aforementioned methods through the following equations. 
For the $\max(\chi_p^2)$ values the significance is calculated through

\begin{equation}
\sigma_{\chi^2} = s^{-1}(0.5 * p_{\rm{value}, \chi_p^2})
\label{eqn:sig_chi2}
.\end{equation}The significance from the $\Delta$TS value is given by

\begin{equation}
\sigma_{\Delta\rm{TS}} = \sqrt{s^{-1}(p_{\rm{value}, \Delta\rm{TS}}, df=1)} \, ,
\label{eqn:sig_TS}
\end{equation}where $s^{-1}$ is the inverse survival function (inverse of $s$).
The comparison of the distribution of significance for each method is shown in the bottom panel of Fig. \ref{fig:sensitivity2}. The middle and bottom panels of Fig.\ref{fig:sensitivity2} include a black dashed line corresponding to unity ($x=y$), serving as a visual reference for assessing agreement between the two distributions.

\section{Conclusion}
There are several methods for determining the energy dependence of a gamma-ray source. Here we have introduced a method in the Gammapy framework that involves the comparison of the spatial model in different energy bins. This method provides a full statistical assessment of the energy dependence, which is often difficult to quantify. 
We show two examples to demonstrate the capabilities of the tool. 
The first example uses H.E.S.S. data towards \hessj, which shows an energy-dependent morphology with a significance of 9.8$\sigma$: the emission is more extended at lower energies and is smaller near the pulsar position at higher energies. The second example examines a weakly extended PWN, modelled using a mock dataset, revealing an energy-dependent morphology with a significance of 9.7$\sigma$ despite the small spatial scale approaching the instrument’s PSF. 
This is an important advancement as it will allow us to resolve the gamma-ray morphology in different energy bands in detail with the upcoming CTAO. With its improved angular resolution and sensitivity, and broader energy coverage compared to current instruments, CTAO will provide important insights into the structure of these sources. 
Morphology studies become more feasible with CTAO's improved PSF, allowing the study of sources with extensions close to the PSF size.

We assessed the accuracy of our technique by conducting a number of tests on a mock dataset that exhibits an energy-dependent morphology. We utilised 1000 simulations to ensure the method is robust. 
In general, we find that these results are in good agreement with those derived from established statistical equations and methods. 

\begin{acknowledgements}
We thank the H.E.S.S. collaboration for allowing us to employ the data used in this publication. 
This paper describes a tool within Gammapy, which is a core Python package in TeV gamma-ray astronomy. We acknowledge the Gammapy development team and any users which provided feedback. In addition, we also acknowledge Astropy \citep[a Python package with which we are affiliated with,][]{2013_astropy}, and other important Python packages we depend on: Numpy, Scipy and Matplotlib.
\end{acknowledgements}

\bibliographystyle{aa}
\bibliography{bibliography}

\begin{appendix} 

\section{Code example output}

\begin{figure}[hbt]
\centering
\includegraphics[width=\columnwidth]{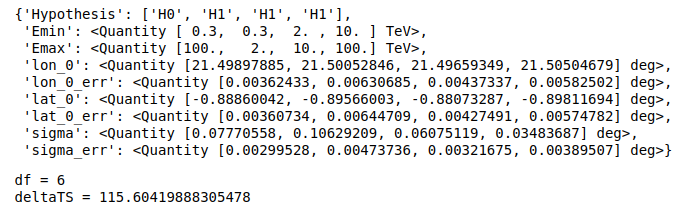}
\caption{Output from the code example shown in Fig. \ref{fig:code_snippet}.}
\label{fig:code_output}
\end{figure}

\section{Chi-squared maximum derivations}
\label{sec:calc_chi2_max}
For a single chi-squared measurement,\begin{equation}
P(x>X) = \int_X^\infty  f(x, df)
\end{equation}where $f(x, df)$ is the chi-squared PDF and $df$ is the number of degrees of freedom. 
For any $x$ to be lower than $X$, the probability is

\begin{equation}
 P(\forall x_i, x_i<X) = \prod_i  \int_X^\infty f(x, df)
.\end{equation}Therefore, the probability that at least one $x$ is larger than $X$ is
\begin{equation}
\begin{split}
P(\exists i, x_i>X) &= 1 - P(\forall x_i, x_i<X) \\
&= 1 - \left( \int_X^\infty f(x, df) \right)^N
\end{split}
\label{eqn:cdf}
.\end{equation}This is explicitly the CDF. 
To obtain the PDF, we had to differentiate the CDF (Eq. \ref{eqn:cdf}), which led to

\begin{equation}
\begin{split}
P(x, df) &= N f(x, df) \left( \int_X^\infty f(x, df) \right)^{(N-1)} \\
&= N f(x, df) \left( F(x, df) \right) ^{(N-1)}
\end{split}
.\end{equation}

\end{appendix}

\end{document}